\documentclass[aip,rsi, floatfix,showpacs,superscriptaddress]{revtex4-1}
\usepackage{graphicx}
\usepackage{amssymb}
\usepackage{amsmath}
\usepackage{soul}
\usepackage[usenames]{color}
\usepackage{ulem}

\ifx\pdftexversion\undefined
\usepackage[dvips]{hyperref}
\else
\usepackage{hyperref}
\fi

\begin{document}

\title{An oscillator circuit to produce a radio-frequency discharge and application to metastable helium saturated absorption spectroscopy}
\author{F. Moron}
\author{A. L. Hoendervanger}
\author{M. Bonneau}
\author{Q. Bouton}
\author{A. Aspect}
\author{D. Boiron}
\author{D. Cl\'ement}
\email[Author to whom correspondence should be adressed. Electronic mail: ]{david.clement@institutoptique.fr}
\author{C. I. Westbrook}
\affiliation{Laboratoire Charles Fabry, Institut d'Optique, CNRS, Univ Paris Sud, 2 Avenue Augustin Fresnel 91127 PALAISEAU cedex}

\begin{abstract}
We present an rf gas discharge apparatus which provides an atomic frequency reference for laser manipulation of metastable helium. 
We discuss the biasing and operation of a Colpitts oscillator in which the discharge coil is part of the oscillator circuit. 
Radiofrequency radiation is reduced by placing the entire oscillator in a metal enclosure. 
\end{abstract}

\maketitle

\textit{Introduction.}  At present, all laser cooling and trapping of noble gas atoms is performed using atoms in excited, metastable states \cite{Vassen12}.  An important part of such manipulation experiments is an absolute optical frequency reference to maintain the laser frequencies at appropriate values. Such a reference is typically provided by an auxiliary low pressure gas cell in which a plasma discharge produces metastable atoms and in which saturated absorption spectroscopy can provide a spectroscopic signal to lock a laser on a desired atomic transition. 
Such discharge cells have employed both direct current (dc) \cite{Lu96, Lawler81} and radiofrequency (rf) \cite{Chen01, Sukenik02, Hoult92, Koelemeij05} techniques. Radiofrequency discharges are appealing because they lend themselves to the use of sealed glass cells \cite{Sukenik02, Hoult92, Koelemeij05} with no internal metal parts. On the other hand, they have the drawback of radiating rf power into a laboratory in which many other sensitive electrical measurements are being performed, rf evaporation in a magnetic trap for example \cite{Browaeys01}. In addition, when rf power is generated by an external oscillator and amplifier, impedance matching to the discharge coil is important and mismatch can result in substantial losses \cite{Sukenik02}.

In this work, we describe an efficient, low cost apparatus which provides good spectroscopic signals in an rf-discharge cell of helium. To avoid impedance matching problems, we use a Colpitts oscillator design in which the discharge coil is included as part of the oscillator circuit \cite{Clifford47}. 
We describe a biasing method which allows us to easily vary the rf amplitude during operation. This feature is important because the voltage necessary to strike the discharge is much higher than that necessary to maintain it or that which optimizes the saturated absorption signal. The entire apparatus can operate in a metal enclosure, thus limiting rf interference in the laboratory.  
\newline

\begin{figure}[ht]
\includegraphics[width=0.5\columnwidth]{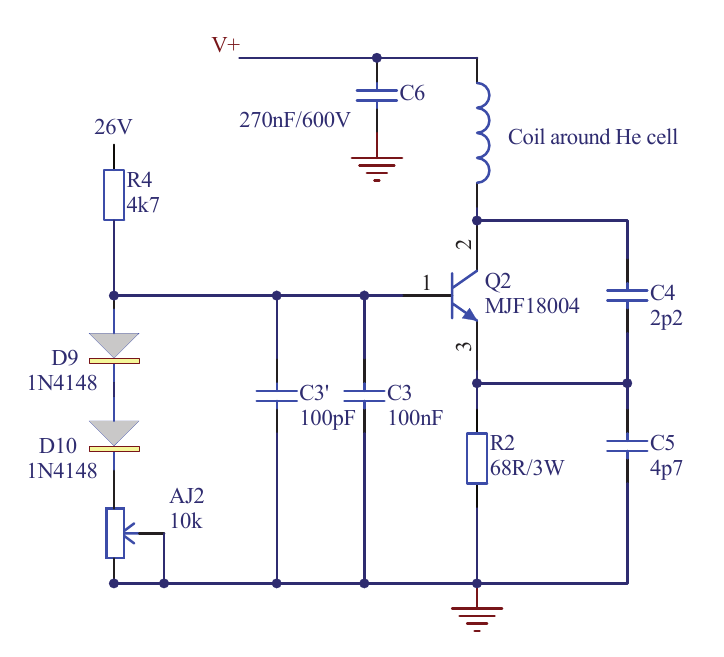} 
\caption{Colpitts oscillator generating an rf signal in the frequency range $12-21~$MHz with varying amplitude $V_{+}$ (160-360 V) in the coil which is wrapped around a sealed glass cell containing about 1~mbar of He gas.}
\label{Fig1}%
\end{figure}

\textit{Radio-frequency Colpitts oscillator.} The oscillator we use to generate the rf signal is a $LC$ oscillator circuit in the Colpitts configuration \cite{Clifford47} (see Fig.~\ref{Fig1}). The resonant frequency $f_{0}$ is close to $21~$MHz before the plasma turns on in the cell. Our approach shares features with the work of Ref.\cite{May86} where the $LC$ oscillator circuit is in the Hartley configuration. In both circuits, the resonant frequency is set by the inductance of the coil wrapped around the glass cell (as well as by the capacitors $C4$ and $C5$ of Fig.~\ref{Fig1}), therefore avoiding any problem of impedance matching between the oscillator and the load \cite{Sukenik02}. Below we discuss our circuit in more detail and emphasize the differences with the work of May et al. \cite{May86}, especially with regard to the possibility to start the plasma without the need for an additional Tesla coil.
 
\begin{figure*}[ptb]
\includegraphics[width=\columnwidth]{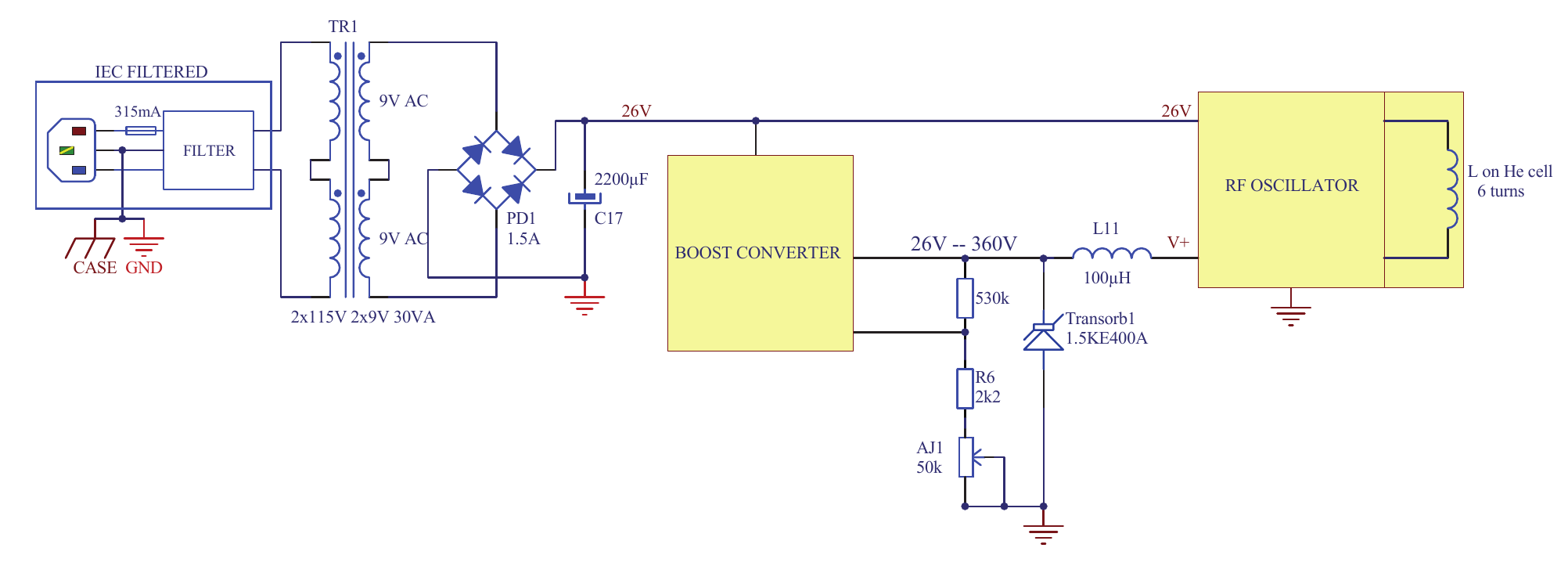} 
\caption{Schematic of the electronic circuit located inside the box (see details in appendix). 
The potentiometer $AJ1$ controls the voltage $V_{+}$ and can be seen on the side of the box in the picture of Fig.~\ref{Fig3}.}%
\label{Fig2}%
\end{figure*}

Starting a plasma discharge requires much more power than maintaining one. For this reason, in addition to the supply which generates the steady state power, a Tesla coil is often used to strike the discharge \cite{May86}. To avoid the Tesla coil, we supply the Colpitts circuit with a varying voltage $V_{+}$ allowing a high enough power to start the plasma (up to 30 W with $V_{+}=360~$V). $V_{+}$ is given by a boost converter the details of which can be found later in the text (see Fig.~\ref{Fig2}). Once the plasma is on, we reduce $V_{+}$  to a value which optimizes our spectroscopic observations (see text below and Fig.~\ref{Fig4}). The need to tune $V_{+}$ made us choose the scheme described in Fig.~\ref{Fig1} where the base bias of the transistor $Q2$ is connected to a constant positive voltage of 26 V, unlike common base Colpitts oscillators. This solution has two advantages. First, it allows us to solve the problem of adjusting the circuit to obtain an oscillation. Indeed, oscillation of the Colpitts circuit is critical \cite{Clifford47} and requires adjustment of the potentiometer $AJ2$ (see Fig.~\ref{Fig1}) to adapt the transistor base current. In the common base Colpitts configuration, the tuning of $V_{+}$ would require adjusting $AJ2$ every time $V_{+}$ is modified. Having a constant base bias voltage (and thus a constant transistor base current) allows a single adjustment of $AJ2$. Second, electrical consumption is much reduced with the constant 26 V voltage compared to that using $V_{+}$.

Because the inductance of the circuit depends on the state of the plasma, the resonant frequency of the circuit changes with the applied voltage $V_{+}$. In our circuit, the resonant frequency with the plasma on varies from $\sim12~$MHz at $V_{+}=140~$V to $\sim20~$MHz at $V_{+}=360~$V. In practice, we start the plasma discharge at 21 MHz with full rf power ($\simeq 30~$W with $V_{+}=360~$V) and then we reduce the rf amplitude to a working point where the resonant frequency happens to be 14 MHz ($\simeq 7~$W with $V_{+}=170~$V).
\newline

\textit{Electronics inside the closed box.}
The oscillator is supplied by a boost converter as depicted in Fig.~\ref{Fig2} (see Fig.~\ref{Fig5} for a more detailed diagram). The use of a boost converter is motivated by its efficient power production and its compactness. The converter provides enough power to start the plasma inside the cell and it allows us to later diminish the electrical power (through the potentiometer $AJ1$) in order to control the metastable atom density in the cell and thus the amount of absorption of the laser light propagating through the cell (see Fig.~\ref{Fig4}).

Two frequency filtering elements have been placed in the electrical circuit. An inductance of 100$~\mu$H (inductance L11 in Fig.~\ref{Fig2}) decouples the Colpitts oscillator from the boost converter. As the box is directly plugged onto the electrical network at 50~Hz, a filter (IEC filter) prohibits any leak rf signal to propagate in the electrical network.

\begin{figure}[ht]
\includegraphics[width=0.5\columnwidth]{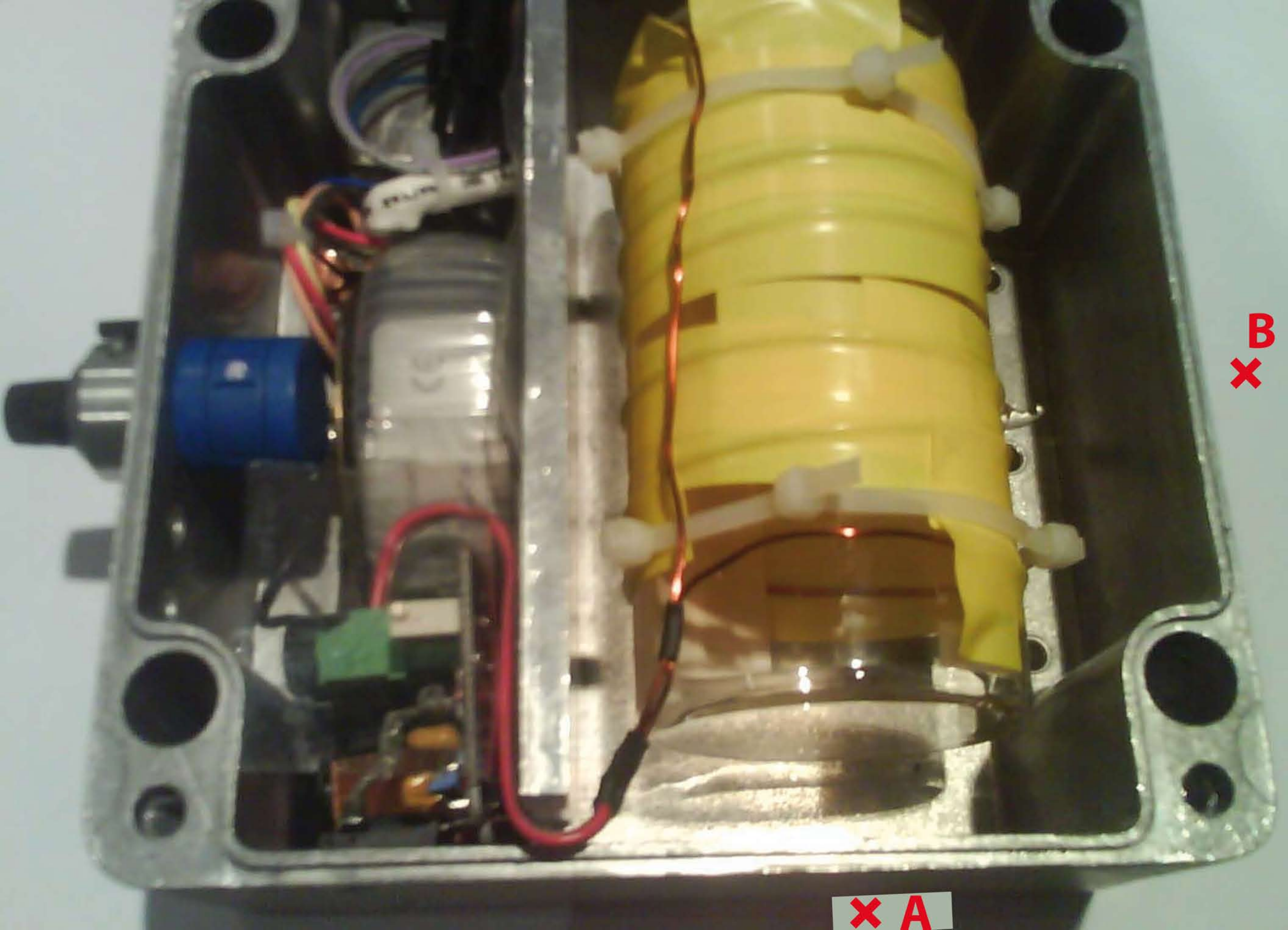} 
\caption{View inside the box with the rf-discharge cell. The box has dimensions $160 \times 160 \times 90~$mm. A metal shield separates the transformer and the electronic circuit from the glass cell and the coils. Points A and B are the positions where we measure the current induced by the radiated magnetic field in a 5~cm-loop. Point A lies in front of the 5mm hole  to allow the laser beam to enter the cell.}%
\label{Fig3}%
\end{figure}

\begin{figure}[ht]
\includegraphics[width=.5\columnwidth]{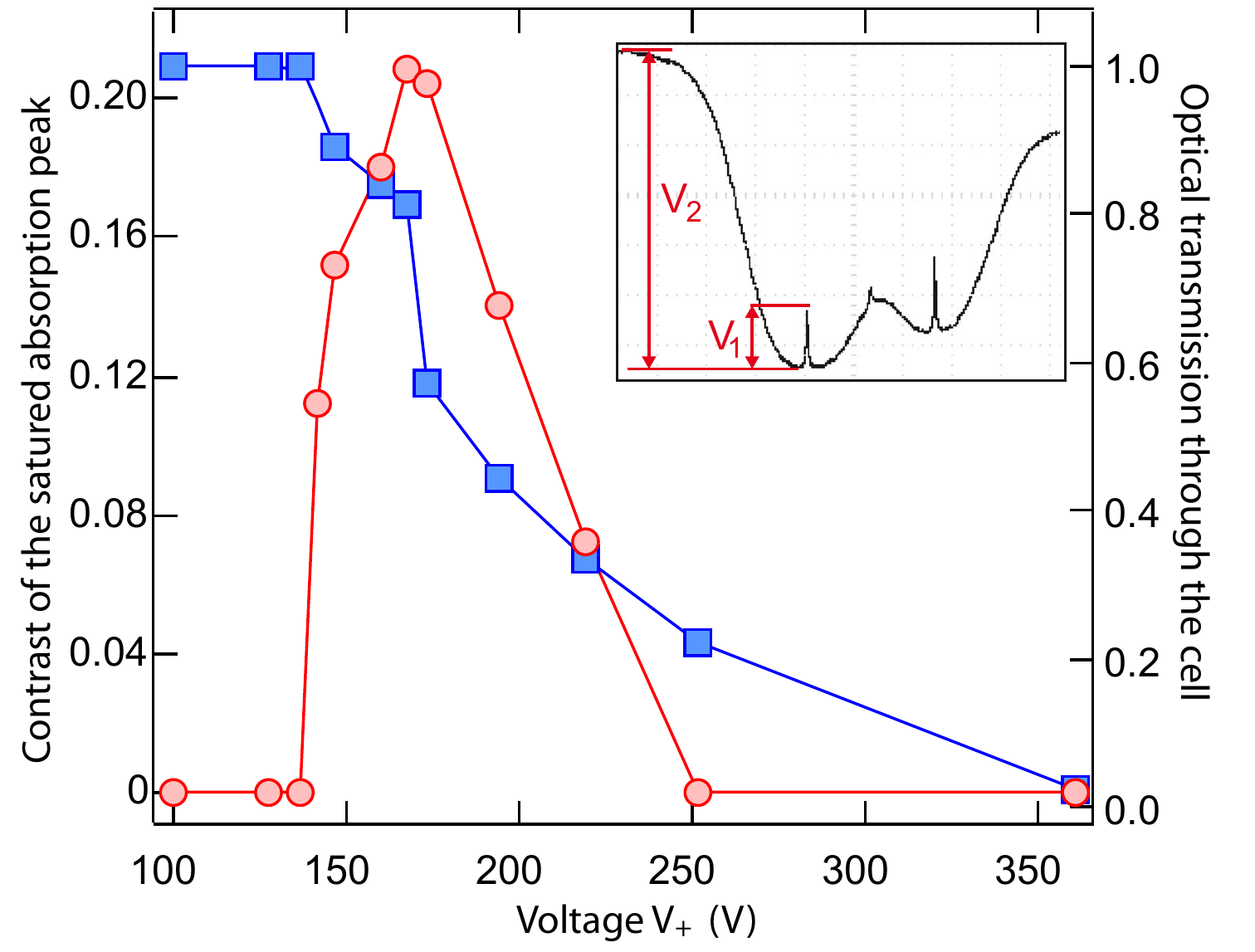} 
\caption{Contrast of the saturated absorption peak (red circles, left scale) and optical transmission of the laser through the glass cell (blue squares, right scale) as a function of the voltage $V_{+}$. 
The minimum voltage to maintain the plasma is $V_{+}=130~$V. Data points below this value are taken without plasma on. Lines are guides to the eyes. Inset: picture of the saturated absorption spectrum of metastable Helium. The ratio of voltages $V_{1}/V_{2}$ defines the contrast of the saturated absorption peak of the $2^3$S$_{1}-2^3$P$_{2}$ transition. The other absorption peak visible corresponds to the transition $2^3$S$_{1}-2^3$P$_{1}$}%
\label{Fig4}%
\end{figure}

\begin{figure*}[ht]
\includegraphics[width=.95 \columnwidth]{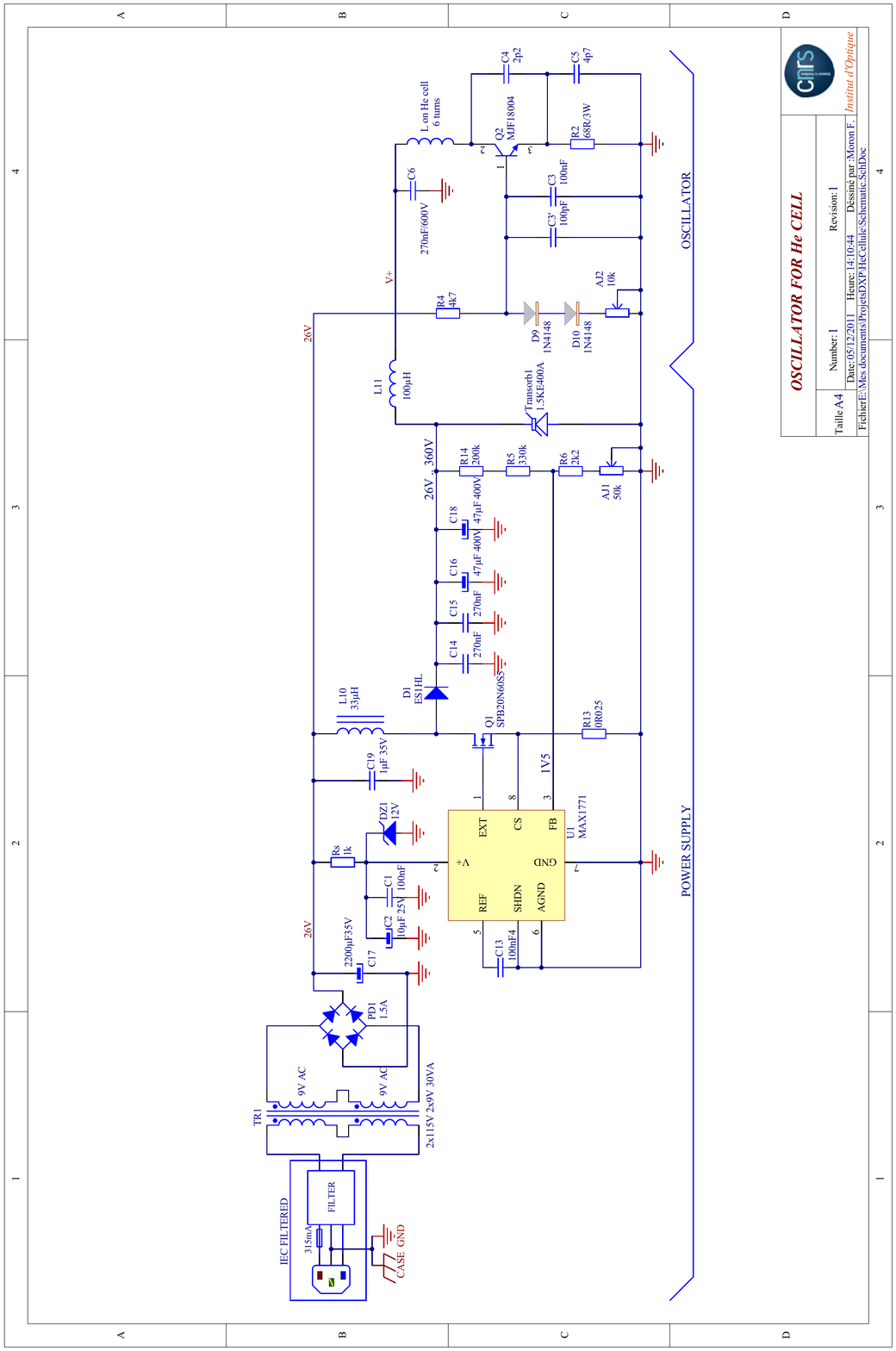} 
\caption{Complete schematic diagram of the electronics.}%
\label{Fig5}%
\end{figure*}

All components shown in Fig.~\ref{Fig2} are placed in an aluminum box of thickness 3 mm to reduce the rf signal radiated into the laboratory. For such a box, the skin depth for a 10 MHz rf signal is 35 $\mu$m, corresponding to an enormous attenuation over 3mm. However, the holes (of diameter 5~mm) required for letting the laser beam propagate through the cell will limit the attenuation of the radiated rf signal in the laboratory. The picture Fig.~\ref{Fig3} shows a view inside the box. The electrical circuit of Fig.~\ref{Fig2} is on the left-hand side and the He glass cell wrapped with the coil on the right-hand side of the picture. Having rf signals confined inside the box avoids both problems of rf radiation from the coil wrapped around the cell as well as rf radiation through cables that would  connect an external rf generator \cite{Sukenik02}.

Ignition of the plasma is more difficult with the box closed, presumably because of parasitic capacitive coupling to the box walls.  
Indeed, the initial version of the boost converter did not provide enough power (maximum value $V_{+}\simeq280~$V instead of 360~V) to start the discharge with the box closed.  Using a more powerful boost converter as described here, permits ignition of the plasma with the box closed in a few seconds. We emphasize that the conditions for ignition of the plasma are related to the properties of the coil, the box and of the sealed glass cell ({\it e.g.} the gas pressure, about 1 mBar in our case). 
\newline

\textit{Measurement of rf attenuation.} 
To test the attenuation of the radiated rf signal when the box is closed, we have measured the voltage amplitude of the signal picked up by a loop of copper wire with a diameter of 5~cm similar to that of the glass cell. The induced current in the loop is proportional to the rf magnetic field and the measured voltage amplitude is therefore proportional to the amplitude of the radiated rf field. We have repeated this measurement at point A, in front of the hole made in the box (just in front of the glass cell) and at point B, on the side of the glass cell, both with the box closed and opened. The attenuation of the magnetic field amplitude at point A is $\sim100$ while it is $\sim300$ at point B. This difference is expected as point A lies in front of the small hole made in the box for the laser beam to propagate. In this respect point B is more representative of the attenuation on the radiated rf magnetic field in the lab room.
\newline

\textit{Absorption spectroscopy signal.} 

We have used our device to lock a laser to the atomic transition $2^3$S$_{1}-2^3$P$_{2}$ at $1083~$nm of metastable He. Our laser source is a fiber-laser with 2W output power. A small part of the laser power is double-passed in an acousto-optic modulator to produce a 10 kHz frequency modulation (FM) of the laser frequency. The FM laser beam is sent through the cell in a saturated absorption spectroscopy configuration. 

When the plasma is on, optical transmission goes to zero with increasing $V_{+}$ as the density of metastable Helium atoms increases. In Fig.~\ref{Fig4}, we plot a measurement of the optical transmission through the glass cell as a function of the voltage $V_{+}$. The transmission is unity for voltages below the minimum for maintaining the plasma (130~V). 
To choose the working point we monitor the amplitude of the saturated absorption peak. The contrast of this peak, plotted in Fig.~\ref{Fig4}, is defined as the ratio of the voltage amplitude of the saturated absorption peak to that of the total absorption (as sketch in the inset of Fig.~\ref{Fig4}). 
The optimum value for $V_{+}$ corresponds to the deepest saturated absorption feature. 
It happens to be $V_{+}\simeq 170~$V, well below the ignition point of the plasma and just above the minimum voltage to maintain the plasma. Finally an error signal is obtained from the demodulated absorption signal \cite{Vansteenkiste91}. We easily obtain stable locking with a few mW of laser power delivered to the cell. Thanks to the use of a toroidal transformer (TR1 in Fig.~\ref{Fig2}), we observe no 50 Hz oscillations in the absorption signal. 

\textit{Conclusion.} 
With the exception of the glass cell, all the components of our device are inexpensive and readily available. 
The attenuation provided by the metal box seems to be adequate for continuous use in our laboratory.
The design should be easily adaptable to other metastable species. 

\begin{acknowledgments}
We acknowledge support from the Triangle de la Physique - contract 2010-062T, the IFRAF Institute, the ANR and the ERC - Grant 267 775 Quantatop.
\end{acknowledgments}

\begin{thebibliography}{21}
\expandafter\ifx\csname natexlab\endcsname\relax\def\natexlab#1{#1}\fi
\expandafter\ifx\csname bibnamefont\endcsname\relax
  \def\bibnamefont#1{#1}\fi
\expandafter\ifx\csname bibfnamefont\endcsname\relax
  \def\bibfnamefont#1{#1}\fi
\expandafter\ifx\csname citenamefont\endcsname\relax
  \def\citenamefont#1{#1}\fi
\expandafter\ifx\csname url\endcsname\relax
  \def\url#1{\texttt{#1}}\fi
\expandafter\ifx\csname urlprefix\endcsname\relax\def\urlprefix{URL }\fi
\providecommand{\bibinfo}[2]{#2}
\providecommand{\eprint}[2][]{\url{#2}}

\bibitem{Vassen12} W. Vassen {\it et al.}, to be published Rev. Mod. Phys., {\bf 84}, (2012).

\bibitem{Lu96} W. Lu {\it et al.}, Rev. Sci. Instrum. {\bf 67}, 3003 (1996).

\bibitem{Lawler81} J. E. Lawler {\it et al.}, J. Appl. Phys. {\bf 52} 4375 (1981).

\bibitem{Chen01} C. Y. Chen {\it et al.}, Rev. Sci. Instrum. {\bf 71}, 271 (2001).

\bibitem{Sukenik02} C. I. Sukenik and H. C. Busch, Rev. Sci. Instrum. {\bf 73}, 493 (2002).

\bibitem{Hoult92} D. I. Hoult and C. M. Preston, Rev. Sci. Instrum. {\bf 63}, 1927 (1992).

\bibitem{Koelemeij05} J. C. J. Koelemeij, W. Hogervorst and W. Vassen, Rev. Sci. Instrum. {\bf 76}, 033104 (2005).

\bibitem{Browaeys01} A. Browaeys et al., Phys. Rev. A {\bf 64} 034703 (2001).

\bibitem{Clifford47} A. E. Clifford et al., "Electronic circuits and tubes", McGraw-Hill Book Co., New-York (1947)

\bibitem{May86} R. D. May and P. H. May, Rev. Sci. Instrum. {\bf 57}, 2242 (1986).

\bibitem{Ding06} Y. Ding et al., Rev. Sci. Instrum. {\bf 77}, 126105 (2006).

\bibitem{Vansteenkiste91} N. Vansteenkiste et al. J. Phys. II France {\bf 1} 1407-1428 (1991).

\end{thebibliography}
\end{document}